\documentclass[OSA,JOSAB,twocolumn]{revtex4}
\usepackage{amsmath}
\usepackage{amssymb}
\usepackage{graphics}
\begin{document}
\title{Truncated Sum Rules and their use in Calculating Fundamental Limits of Nonlinear Susceptibilities} 
\author{Mark G. Kuzyk}
\address{Department of Physics and Astronomy, Washington State University, Pullman,
Washington  99164-2814 \\ email: kuz@wsu.edu}
\newcommand{\bibs}{archive}
\date{\today}
\begin{abstract}
Truncated sum rules have been used to calculate the fundamental limits of the nonlinear susceptibilities; and, the results have been consistent with all measured molecules.  However, given that finite-state models result in inconsistencies in the sum rules, it is not clear why the method works.  In this paper, the assumptions inherent in the truncation process are discussed and arguments based on physical grounds are presented in support of using truncated sum rules in calculating fundamental limits.  The clipped harmonic oscillator is used as an illustration of how the validity of truncation can be tested; and, several limiting cases are discussed as examples of the nuances inherent in the method.
\end{abstract}
\maketitle

\section{Introduction}
Truncated sum rules were first used at the turn of the century to calculate the fundamental limits of the off-resonant hyperpolarizability $\beta$,\cite{kuzyk00.01,kuzyk03.02} and the second hyperpolarizability, $\gamma$.\cite{kuzyk00.02,kuzyk03.01} The role of off-diagonal elements, such as measured with Hyper Rayleigh Scattering, were also considered.\cite{kuzyk01.01}  These calculations were later applied to calculating the resonant two-photon absorption cross-section\cite{kuzyk03.03} as well as the maximum possible two-photon absorption cross-section when a molecule is doubly-resonant with the excitation source.\cite{kuzyk04.02}  The theory is supported by the fact that no molecule has ever been found to break the fundamental limit.\cite{Kuzyk03.04} 

While all molecules ever measured were found to fall below the limit, it was pointed out that there is a gap between the fundamental limit and the molecules with the largest measured nonlinear-optical susceptibility.\cite{kuzyk03.02,Kuzyk03.05,kuzyk03.01}  This apparent limit, a factor of $10^{3/2}$ below the fundamental limit, had two possible implications.  Either, the theory could be flawed so that it overestimated the fundamental limit by more than a factor of thirty; or there was another factor that acted to suppress the nonlinear-optical response.  To test the hypothesis that the theory has overestimated the nonlinear response, all one needs to do is show that a system exists whose hyperpolarizability exceeds the apparent limit.  The clipped harmonic oscillator, whose hyperpolarizability can be calculated analytically without approximation, has a hyperpolarizability that is an order of magnitude larger than the apparent limit, yet about a factor of two below the fundamental limit.\cite{Tripa04.01}  Given that very talented organic chemists have been working almost three decades to improve the hyperpolarizability of molecules through structure-property studies, and no molecule has ever been found to breach the apparent limit, it is probably prudent to conclude that the best organic molecules fall below the apparent limit.

While nanoengineering methods have been used to increase the second hyperpolarizability per molecule to within a factor of two of the fundamental limit,\cite{wang04.01} a more careful analysis of this case shows that the interactions between molecules makes them respond collectively.  As such, the collection of molecules is acting like a single supermolecule where its response is still well below the apparent limit.  So, the state of affairs with regards to molecular hyperpolarizabilities is that all molecules ever measured and all analytical calculations of $\beta$ and $\gamma$ ever performed fall below the fundamental limit, though some of the calculations come close.  This set of evidence supports the theory of fundamental limits.  So, while the process of truncating the sum rules and applying them to the analysis of the Sum Over States (SOS) expression for $\beta$ could lead to pathologies, the results of this process seems consistent with observation.

\section{Theory of the Fundamental Limit}

This section develops the theory of the fundamental limits of the hyperpolarizability (the second hyperpolarizability follows along the same lines but is algebraically more messy, so will not be presented here).  The method presented follows the literature,\cite{kuzyk00.01,kuzyk03.02} but, close attention will be focused on the assumptions underlying the calculations and all of the details will be presented with commentary on their meaning.  

We being from the sum rules,
\begin{equation}
\sum_{n=0}^{\infty} \left( E_n - \frac {1} {2} \left( E_m + E_p \right) \right) x_{mn} x_{np} = \frac {\hbar^2 N} {2m} \delta_{m,p},
\label{sumrule}
\end{equation}
which relate the matrix elements of the position operator, $x_{nm}$, to the energy differences, $E_{nm} = E_n - E_m$, between the states $n$, and $m$.  Since the sum rules derive directly from the Schr\"{o}dinger Equation, they are of a fundamental nature and apply to all systems in which one can write a potential energy function.  Note that Equation~\ref{sumrule} yields a distinct equation for each distinct pair of values of $m$, and $p$.  As such, we will refer to the sum rule equation with indices $m$ and $p$ as Equation $X_{mp}$.  When the sum rules are truncated to include only $\ell$ terms (i.e. the sum over $n$ in Equation \ref{sumrule} ranges from 0 to $\ell-1$), we will refer to the truncated sum rule as $x_{nm}^{(\ell)}$.

\subsection{Linear Susceptibility}

We begin by treating the linear susceptibility, which is the simplest possible case. A quantum perturbation calculation of the polarizability of a one-dimensional system, $\alpha$, yields,
\begin{equation}
\alpha = \frac {e^2} {\hbar} \sum_{n=1}^{\infty} \left| x_{n0} \right|^2 \left[ \frac {1} {\omega_{n0} - \omega - i \gamma_{n0}} + \frac {1} {\omega_{n0} + \omega - i \gamma_{n0}} \right] , \label{alpha}
\end{equation}
where $\omega$ is the frequency of light, $\omega_{n0}$ the transition frequency $\omega_n -\omega_0$ between states $n$ and $0$, $\gamma_{n0}$ the width of the transition, and $e$ the electron charge.  We consider the off-resonance limit, where $\omega \rightarrow 0$ and calculate the real part of the polarizability, where $Re[\alpha]$, where $\gamma_{n0}=0$,
\begin{equation}
\alpha = 2 e^2 \sum_{n=1}^{\infty} \frac {\left| x_{n0} \right|^2} {E_{n0}}  , \label{alphaRealNonResonant}
\end{equation}
where we have used $E_{n0} = \hbar \omega_{n0}$.

Now we use the non-truncated sum rules to simplify Equation~\ref{alphaRealNonResonant}.  The sum rule Equation $X_{00}$ (Equation~\ref{sumrule} with $m=p=0$) is given by
\begin{equation}
\sum_{n=0}^{\infty}  E_{n0} \left| x_{n0} \right|^2 = \frac {\hbar^2 N} {2m} .
\label{groundsumrule}
\end{equation}
To use $X_{00}$, we rewrite Equation~\ref{alphaRealNonResonant} as follows,
\begin{equation}
\alpha = 2 e^2 \sum_{n=1}^{\infty} \frac {E_{n0} \left| x_{n0} \right|^2} {E_{n0}^2} \leq 2 e^2 \sum_{n=1}^{\infty} \frac {E_{n0} \left| x_{n0} \right|^2} {E_{10}^2}, \label{alphaRealNonResonant2}
\end{equation}
where the inequality follows from the fact that $E_{n0} \leq E_{10}$ and all the terms in the sum are positive definite.  Substituting Equation~\ref{groundsumrule} into Equation~\ref{alphaRealNonResonant2}, we get
\begin{equation}
\alpha \leq  \left( \frac{e^2 \hbar^2} {m} \right) \frac {N} {E_{10}^2} \equiv \alpha_{MAX} . \label{alphaMAXoffRes}
\end{equation}
Notice that if we write sum rule Equation $X_{00}$ as,
\begin{equation}
\left| x_{10} \right|^2 = \frac {\hbar^2} {2mE_{10} } N - \sum_{n=2}^{\infty}  \frac {E_{n0}} {E_{10}} \left| x_{n0} \right|^2,
\label{groundsumrule2}
\end{equation}
then clearly, the maximum value of $\alpha$ is obtained by placing all of the oscillator strength into the first excited state and setting all of the other transition moments to zero, that is $x_{n0}=0$ for $n>1$.  Note that when all of the transition moment is concentrated into the first excited state, $x_{10}$ takes on its maximum value as given by Equation~\ref{groundsumrule2}.

The important lesson of this calculation is that in the limit of maximal $\alpha$, the oscillator strength gets concentrated into the transition between the ground and first excited state.  It is instructive to treat this problem as a three-level system where the SOS expression includes only the ground and first two excited states; and similarly, to truncate the sum rules to three levels.

The three-level model for $\alpha$ is given by
\begin{equation}
\alpha_{3L} = 2e^2  \left\{ \frac {\left| x_{10} \right|^2} {E_{10}} + \frac {\left| x_{20} \right|^2} {E_{20}} \right\}, \label{threelevelalpha}
\end{equation}
where by definition, state $2$ is of higher energy than state $1$.  Using the three-level sum rule given by Equation $X_{00}^{(3)}$ and including only three states (0, 1, and 2) to eliminate $x_{02}$, Equation (\ref{threelevelalpha}) becomes:
\begin{equation}
\alpha_{3L}  = 2e^2 \frac {\left| x_{10} \right|^2} {E_{10}} \left\{ 1 + \frac {N \hbar^2} {2m E_{20}} \left( \frac {E_{10}} {E_{20}} \right) \frac {1} {\left| x_{10} \right|^2} - \left( \frac {E_{10}} {E_{20}} \right)^2 \right\}. \label{threelevelalpha2}
\end{equation}
Equation (\ref{threelevelalpha2}) can be rewritten in the form,
\begin{equation}
\alpha_{3L} = \alpha_{MAX} \left( X^2 + E^2 - E^2 X^2 \right) \equiv  \alpha_{MAX} h(X,E), \label{alphaform}
\end{equation}
where
\begin{equation}
X = \frac {x_{10}} {x_{10}^{MAX}} , \label{Xfrac}
\end{equation}
with
\begin{equation}
x_{10}^{MAX} = \sqrt{ \frac {\hbar^2 N} {2mE_{10}} }, \label{xsum}
\end{equation}
and
\begin{equation}
E = \frac {E_{10}} {E_{20}}. \label{Efrac}
\end{equation}
Note that with the above definitions, $-1 \leq x \leq +1$ and $-1 \leq E \leq +1$.  We call Equation \ref{alphaform} the reduced three-level model since the sum rules have reduced the number of parameters.

\begin{figure}
\centering
\includegraphics{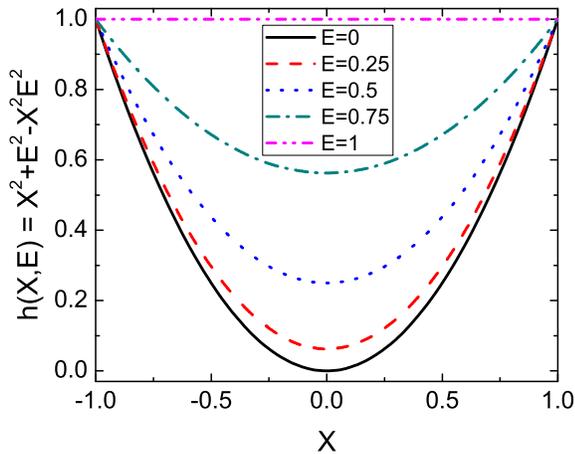}
\caption{A plot of $h(X,E)$.\label{fig:alpha3L}}
\end{figure}

Using the reduced three-level model, we assume that the energy parameter, $E$, and the transition moment parameter, $X$, are independent variables.  Figure \ref{fig:alpha3L} shows a plot of $h(X,E)$ as a function of $X$ for several values of $E$.
Clearly, $h(X=1,E)=1$ for all values of $E$, so, the three-level model yields the same maximum value of $\alpha = \alpha_{MAX}$ as the infinite-state result.   

The reduced three-level model yields insights into the nature of fundamental limits.  First, the fact that $X=1$ at the fundamental limit clearly shows that all of the oscillator strength must reside in the first excited state.  This is consistent with the results of the infinite level model.  Secondly, note that at $E=1$, when the two excited states are degenerate, the polarizability is at the fundamental limit for all possessible values of the transition moment to the first excited state.  This behavior can be understood by noting that when $E_{10} = E_{20}$, the truncated sum rule $X_{00}^{(3)}$ yields
\begin{equation}
\left| x_{10} \right|^2 + \left| x_{20} \right|^2 = \frac {\hbar^2} {2mE_{10} } N,
\label{groundsumrule(3)}
\end{equation}
which when substituted into Equation \ref{threelevelalpha} yields
\begin{equation}
\alpha_{3L} = \frac {2e^2} {E_{10}}  \left\{ \left| x_{10} \right|^2 + \left| x_{20} \right|^2 \right\} = \frac {N e^2 \hbar^2} {mE_{10}^2} = \alpha_{MAX}. \label{threelevelalphaE=1}
\end{equation}
As such, the degenerate case acts like a two-level model.

This simple example shows that the fundamental limit of the polarizability occurs when all of the oscillator strength is concentrated in the transition to the first excited state.  As such, it is sufficient to consider only a reduced three-level model since this is a good approximation for any quantum system that has a polarizability near its maximum value. 

\subsection{Nonlinear Polarizability}

In the case of the hyperpolarizability and higher-order polarizabilities, the sum-over-states expressions are so complicated that it is not possible to calculate the exact result as we did for $\alpha$.  In the linear case, we concluded that the fundamental limit occurs when all the oscillator strength is concentrated into one transition.  However, the nonlinear susceptibilities depend on two types of terms.  So, it would not be unreasonable to expect that the oscillator strength may need to be shared between at least two excited states.

Assuming that a fundamental limit indeed exists, we therefore use the following ansatz:\cite{kuzyk05.01}
\begin{itemize}
\item{{\bf Ansatz}: In the limit when the static nonlinear polarizability of a bound quantum system in its ground state approaches the fundamental limit, the only allowable non-vanishing matrix elements of the dipole operator are described by a 3$\times$3 matrix.}
\end{itemize}
An important consequence of this ansatz is that the sum rules must be truncated, resulting in a finite number of equations of the form $X_{nm}^{(3)}$, with $m<3$ and $n<3$.  The set of truncated sum rules leads to several problems, such as self inconsistencies.  To deal with inconsistencies and decide which set of sum rules are appropriate, we adopt the following guiding principles:
\begin{enumerate}
\item{The set of sum-rule equations must be self-consistent.}
\item{The predictions following from the application of the sum rules must be consistent with experiment.}
\item{When off-resonance, all measurable quantities should be free from divergences.}
\end{enumerate}

Our first priority is to decide on the fundamental set of sum-rule equations.  First, we note that sum rule equation $X_{22}^{(3)}$ yields a nonsensical result, and, it is inconsistent with the sum rule equation $X_{00}^{(3)}$.  Based on {\em Principles 1} and {\em 2}, we eliminate $X_{22}^{(3)}$.  Secondly, we note that the remaining sum-rule equations are overly restrictive since they predict that all molecules should have the same ratio, $E$.  This violates {\em Principle 2} since molecules with a continuous spectrum of $E$ vales are observed.  As such, we must drop at least one additional equation.  To determine which equation to drop, we go through the process of dropping one equation, and analyzing the ramifications of the remaining set.  We find that the only equation that can be dropped, which leaves a set of equations that do not result in divergences in the calculated off-resonant nonlinear susceptibilities (i.e. obeys {\em Principle 3}), is $X_{21}^{(3)}$.\cite{perez04.01}  If we drop an additional equation, then the result one gets is that there is no limit on the susceptibility.  Since experiment strongly suggests that the nonlinear-optical response can not be arbitrarily large, we conclude that the 4 independent equations  $X_{00}^{(3)}$, $X_{10}^{(3)}$, $X_{11}^{(3)}$, and $X_{20}^{(3)}$ form the complete set.  (Note that Equations  $X_{nm}^{(3)}$ and $X_{mn}^{(3)}$ are complex conjugates of each other, and for a real dipole matrix, are identical.)

The process of applying the truncated sum rules to the three-level model of the sum-over-states expression of the hyperpolarizability, $\beta$ and second hyperpolarizability, $\gamma$, is described in the literature.\cite{kuzyk00.01,kuzyk03.01,kuzyk00.02,kuzyk03.02}  For $\beta$, this yields,
\begin{equation}
\beta = 6 \sqrt{\frac {2} {3}} e^3 \frac {\left| x_{10}^{MAX} \right|^3} {E_{10}^2} G(X) f(E) = \beta_0 G(X) f(E) ,\label{betafG}
\end{equation}
where
\begin{equation}
f(E) = (1-E)^{3/2} \left( E^2 + \frac {3} {2} E + 1 \right),\label{DEFf(E)}
\end{equation}
and
\begin{equation}
G(X) = \sqrt[4]{3} X \sqrt{\frac {3} {2} \left( 1 - X^4\right)}.\label{defG(X)}
\end{equation}
As in the linear case, we assume that $E$ and $X$ are independent.  $f(E)$ peaks at $E=0$ and $G(X)$ peaks at $X=\sqrt[-4]{3}$.  The maximum value of each function is unity, so from Equation \ref{betafG}, we get the fundamental limit
\begin{equation}
\beta_{MAX} = \beta_0 f(0) G(\sqrt[-4]{3}) =  \sqrt[4]{3} \left( \frac {e \hbar} {\sqrt{m}} \right)^3 \left[ \frac {N^{3/2}} {E_{10}^{7/2}} \right] , \label{betaMAX3L}
\end{equation}
where we have made use of Equation \ref{xsum}.  A similar procedure for the second hyperpolarizability yields\cite{kuzyk00.01,kuzyk03.01,kuzyk00.02,kuzyk03.02}
\begin{equation}
-\frac {e^4 \hbar^4} {m^2} \left( \frac {N^2} {E_{10}^5} \right) \leq \gamma \leq 4 \frac {e^4 \hbar^4} {m^2} \left( \frac {N^2} {E_{10}^5} \right) . \label{abslimit}
\end{equation}
Note that the fundamental limit of of positive $\gamma$ is four times larger than for negative $\gamma$.

\section{Analysis}

In this section, we develop a method for analyzing the validity of using truncated sum rules by studying how the truncated series compares with the exact results.  We begin by defining the $\kappa$-matrix, which derives from the sum rules given by Equation \ref{sumrule}, 
\begin{equation}
\kappa_{mp}^{(\ell)} = \delta_{m,p} - \frac {m} {\hbar^2 N} \sum_{n=0}^{\ell} \left( E_{nm} + E_{np} \right)  x_{mn} x_{np} .
\label{kappaDef}
\end{equation}
Note that $\kappa$-matrix is a generalization of an idea first proposed by Champagne and Kirtman.\cite{champ05.01} If the sums rules are exactly obeyed, the $\kappa$-matrix vanishes.  Therefore, kappa is calculated for a given system by substituting the known values of the transition moments and energies into Equation \ref{kappaDef}.  Deviations of $\kappa_{mp}^{(\ell)}$ from zero is a sign that either the act of truncation in inaccurate; or that the matrix elements of the dipole operator and energies are incorrect for that particular system.

In order to test the pathologies associated with truncation, we must start with a potential energy function for which the Schr\"{o}dinger Equation can be solved analytically.  This approach ensures that the energies and dipole matrix elements are physically sound and that pathologies or inaccuracies inherent in approximation techniques are avoided.  The clipped harmonic oscillator (CHO) is the ideal system(where the potential is harmonic for $x>0$ and infinite for $x<0$) since it is the simplest case of an asymmetric potential that yields a hyperpolarizability that is near the fundamental limit.\cite{Tripa04.01} The matrix elements of the position operator of the clipped harmonic oscillator (CHO) are given by,
\begin{equation}
x_{mn} = x_{10}^{MAX} g_{mn},
\label{CHO-xmn}
\end{equation}
where the dimensionless matrix $g_{mn}$ is defined by
\begin{equation}
g_{mn} = \frac {2} {\sqrt{\pi}} (-1)^{((m+n)/2)} \cdot \left( \frac {2} {(m-n)^2 - 1 } \right) \cdot \left( \frac {m!! n!!} {\sqrt{m!n!}} \right) ,
\label{CHO-gmn}
\end{equation}
where $m$ and $n$ are restricted to the odd integers.  The energy for state $n$ is given by
\begin{equation}
E_n = \hbar \omega_0 \left( n + \frac {1} {2} \right) .
\label{CHO-energy}
\end{equation}

\begin{figure}
\centering
\includegraphics{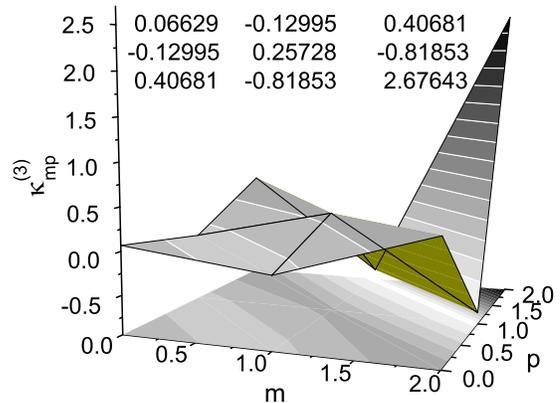}
\caption{Contour plot of $\kappa_{mp}^{(3)}$.  Inset shows matrix elements.\label{fig:kappa3}}
\end{figure}
Figure \ref{fig:kappa3} shows a surface plot of $\kappa_{mp}^{(3)}$ and the inset shows the matrix elements.  Recall that the $\kappa$-matrix is a measure of the fractional deviation of the particular truncated sum-rule equation $X_{nm}^{(\ell)}$.  As such, when truncating to three levels, the sum rule $X_{00}^{(3)}$ with the clipped harmonic oscillator wavefunctions gives a deviation of less than 7\% from the infinite-level case.  The sum rule equations $X_{12}^{(3)}$ and $X_{22}^{(3)}$, which we dropped from our analysis, deviate by 82\% and almost 270\%, respectively.

\begin{figure}
\centering
\includegraphics{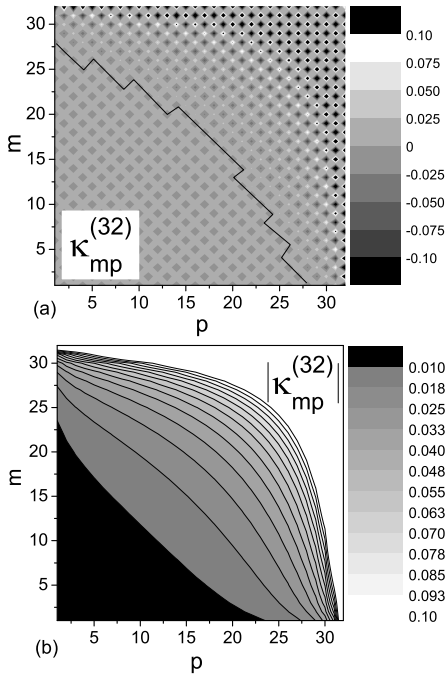}
\caption{Contour plot of (a) $\kappa_{mp}^{(32)}$ and (b) $\left| \kappa_{mp}^{(32)} \right|$.\label{fig:kappa32}}
\end{figure}

Since the clipped harmonic oscillator's hyperpolarizability is near the fundamental limit, we would expect that the three-level sum rules that are kept in the analysis to deviate substantially less than the equations that are dropped.  Figure \ref{fig:kappa32} confirms that the domain of the sum rules used in our maximization calculation is relatively flat.   

To show how quickly the sum rules converge, and how well the exact clipped harmonic oscillator obeys them, we consider a 32-level model.  Figure \ref{fig:kappa32} shows a contour plot of $\kappa_{mp}^{(32)}$ and $\left| \kappa_{mp}^{(32)} \right|$.  First, we note that the deviations alternate sign when $m+p$ changes from even to odd.  (This same pattern is also observed for $\kappa_{mp}^{(3)}$ in Figure \ref{fig:kappa3}.)  The jagged line in Figure \ref{fig:kappa32} represents the point at which $\kappa_{mp}^{(32)} \approx 0.025$.  The areas that are shaded black represented matrix elements of $\kappa$ that are larger than $0.1$.  As such, most of the sum rules are accurate.  Indeed, even $\kappa_{mp}^{(10)}$ converges quickly and the deviations are small for all sum rule equations for $m+p<10$.

Figure \ref{fig:kappa32}b shows the absolute magnitude of $\kappa_{mp}^{(32)}$, which more clearly shows deviations of the $\kappa$-matrix.  In the white region, $\kappa_{mp}^{(32)}>0.1$.  Above the diagonal, from $(m,p) = (1,30)$ to $(m,p) = (30,1)$, $\kappa_{mp}^{(32)}>0.03$.  Recall that in the case of the three-level model, we ignored the two equations: $X_{22}^{(3)}$ for contradicting $X_{00}^{(3)}$ and for being unphysical; and Equation $X_{21}^{(3)}$ since it yielded unphysical results.  The $\kappa$-matrix may be a useful tool in studying which sum rule equations that need to be discarded in higher-level models.

It is interesting to speculate whether the abrupt sign changes in the $\kappa$-matrix may result in a cancellation of terms in the SOS expression for the nonlinear-optical susceptibility; and, if small errors in the matrix elements may lead to large uncertainties in the nonlinear-optical response.

\begin{figure}
\centering
\includegraphics{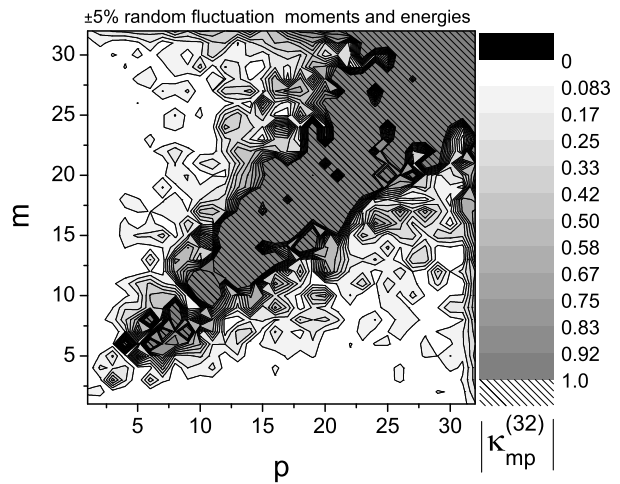}
\caption{Contour plot of $\left| \kappa_{mp}^{(32)} \right|$ where dipole moments and energies randomly are changed by $\pm$5\%.\label{fig:AbsKappa32rand}}
\end{figure}

To test this hypothesis, the matrix elements and energies of the clipped harmonic oscillator were recalculated with random fluctuations of $\pm 5 \, \%$ added to each using the equations,
\begin{equation}\label{eq:XrandomOsc}
x_{nm} \rightarrow x_{nm} \cdot (1+r),
\end{equation}
and
\begin{equation}\label{eq:ErandomOsc}
E_{nm} \rightarrow E_{nm} \cdot (1+r),
\end{equation}
where $r$ is a random number equally distributed in the interval $-0.05 < r <+0.05$.  Figure \ref{fig:AbsKappa32rand} shows a plot of the magnitude of the $\kappa$-matrix for these parameters.  Note that the $\kappa$-matrix is far larger than one would expect if it depended linearly on the random error.  In the white region, $\kappa_{mp}^{(32)}>0.083$, while in many regions, $\kappa_{mp}^{(32)}>1$.  So, it appears that a small error in the matrix elements and energies can lead to a large violation of the sum rules.

\section{Conclusion}

In conclusion, we have shown that the sum rules can be used to calculate the fundamental limits of the nonlinear-optical susceptibility provided that the system approaches a three-level model as $\beta$ and $\gamma$ approach the fundamental limit.  This process of truncating the sum rules and demanding that the fundamental limit occurs in the three-level limit is motivated by exact results that are calculated for the polarizability.  We introduce three guiding principles that aid us in determining which sum rule equations should be dropped, and show that the fundamental limit is unique.  Furthermore, we introduce a method for testing the validity of the transition moments and energy levels and find that even small errors in them can lead to large violations of the sum rules.  As such, this method is ideal for testing the results of semi-empirical calculations.

\section{Acknowledgements}
I thank The National Science Foundation (ECS-0354736) and Wright-Paterson Air Force Base for generously supporting this work.


\begin{thebibliography}{14}
\expandafter\ifx\csname natexlab\endcsname\relax\def\natexlab#1{#1}\fi
\expandafter\ifx\csname bibnamefont\endcsname\relax
  \def\bibnamefont#1{#1}\fi
\expandafter\ifx\csname bibfnamefont\endcsname\relax
  \def\bibfnamefont#1{#1}\fi
\expandafter\ifx\csname citenamefont\endcsname\relax
  \def\citenamefont#1{#1}\fi
\expandafter\ifx\csname url\endcsname\relax
  \def\url#1{\texttt{#1}}\fi
\expandafter\ifx\csname urlprefix\endcsname\relax\def\urlprefix{URL }\fi
\providecommand{\bibinfo}[2]{#2}
\providecommand{\eprint}[2][]{\url{#2}}

\bibitem[{\citenamefont{Kuzyk}(2000{\natexlab{a}})}]{kuzyk00.01}
\bibinfo{author}{\bibfnamefont{M.~G.} \bibnamefont{Kuzyk}},
  \bibinfo{journal}{Phys. Rev. Lett.} \textbf{\bibinfo{volume}{85}},
  \bibinfo{pages}{1218} (\bibinfo{year}{2000}{\natexlab{a}}).

\bibitem[{\citenamefont{Kuzyk}(2003{\natexlab{a}})}]{kuzyk03.02}
\bibinfo{author}{\bibfnamefont{M.~G.} \bibnamefont{Kuzyk}},
  \bibinfo{journal}{Phys. Rev. Lett.} \textbf{\bibinfo{volume}{90}},
  \bibinfo{pages}{039902} (\bibinfo{year}{2003}{\natexlab{a}}).

\bibitem[{\citenamefont{Kuzyk}(2000{\natexlab{b}})}]{kuzyk00.02}
\bibinfo{author}{\bibfnamefont{M.~G.} \bibnamefont{Kuzyk}},
  \bibinfo{journal}{Opt. Lett.} \textbf{\bibinfo{volume}{25}},
  \bibinfo{pages}{1183} (\bibinfo{year}{2000}{\natexlab{b}}).

\bibitem[{\citenamefont{Kuzyk}(2003{\natexlab{b}})}]{kuzyk03.01}
\bibinfo{author}{\bibfnamefont{M.~G.} \bibnamefont{Kuzyk}},
  \bibinfo{journal}{Opt. Lett.} \textbf{\bibinfo{volume}{28}},
  \bibinfo{pages}{135} (\bibinfo{year}{2003}{\natexlab{b}}).

\bibitem[{\citenamefont{Kuzyk}(2001)}]{kuzyk01.01}
\bibinfo{author}{\bibfnamefont{M.~G.} \bibnamefont{Kuzyk}},
  \bibinfo{journal}{IEEE Journal on Selected Topics in Quantum Electronics}
  \textbf{\bibinfo{volume}{7}}, \bibinfo{pages}{774 } (\bibinfo{year}{2001}).

\bibitem[{\citenamefont{Kuzyk}(2003{\natexlab{c}})}]{kuzyk03.03}
\bibinfo{author}{\bibfnamefont{M.~G.} \bibnamefont{Kuzyk}},
  \bibinfo{journal}{J. Chem Phys.} \textbf{\bibinfo{volume}{119}}
  (\bibinfo{year}{2003}{\natexlab{c}}).

\bibitem[{\citenamefont{Kuzyk}(2004)}]{kuzyk04.02}
\bibinfo{author}{\bibfnamefont{M.~G.} \bibnamefont{Kuzyk}},
  \bibinfo{journal}{J. Nonl. Opt. Phys. \& Mat.} \textbf{\bibinfo{volume}{13}},
  \bibinfo{pages}{461} (\bibinfo{year}{2004}).

\bibitem[{\citenamefont{Kuzyk}(2003{\natexlab{d}})}]{Kuzyk03.04}
\bibinfo{author}{\bibfnamefont{M.~G.} \bibnamefont{Kuzyk}},
  \bibinfo{journal}{IEEE Circuits and Devices Magazine}
  \textbf{\bibinfo{volume}{19}}, \bibinfo{pages}{8}
  (\bibinfo{year}{2003}{\natexlab{d}}).

\bibitem[{\citenamefont{Kuzyk}(2003{\natexlab{e}})}]{Kuzyk03.05}
\bibinfo{author}{\bibfnamefont{M.~G.} \bibnamefont{Kuzyk}},
  \bibinfo{journal}{Optics \& Photonics News} \textbf{\bibinfo{volume}{14}},
  \bibinfo{pages}{26} (\bibinfo{year}{2003}{\natexlab{e}}).

\bibitem[{\citenamefont{Tripathi et~al.}(2004)\citenamefont{Tripathi, Moreno,
  Kuzyk, Coe, Clays, and Kelley}}]{Tripa04.01}
\bibinfo{author}{\bibfnamefont{K.}~\bibnamefont{Tripathi}},
  \bibinfo{author}{\bibfnamefont{P.}~\bibnamefont{Moreno}},
  \bibinfo{author}{\bibfnamefont{M.~G.} \bibnamefont{Kuzyk}},
  \bibinfo{author}{\bibfnamefont{B.~J.} \bibnamefont{Coe}},
  \bibinfo{author}{\bibfnamefont{K.}~\bibnamefont{Clays}}, \bibnamefont{and}
  \bibinfo{author}{\bibfnamefont{A.~M.} \bibnamefont{Kelley}},
  \bibinfo{journal}{J. Chem. Phys.} \textbf{\bibinfo{volume}{121}},
  \bibinfo{pages}{7932} (\bibinfo{year}{2004}).

\bibitem[{\citenamefont{Chen et~al.}(2004)\citenamefont{Chen, Kuang, Wang, and
  Sargent}}]{wang04.01}
\bibinfo{author}{\bibfnamefont{Q.~Y.} \bibnamefont{Chen}},
  \bibinfo{author}{\bibfnamefont{L.}~\bibnamefont{Kuang}},
  \bibinfo{author}{\bibfnamefont{Z.~Y.} \bibnamefont{Wang}}, \bibnamefont{and}
  \bibinfo{author}{\bibfnamefont{E.~H.} \bibnamefont{Sargent}},
  \bibinfo{journal}{Nano. Lett.} \textbf{\bibinfo{volume}{4}},
  \bibinfo{pages}{1673} (\bibinfo{year}{2004}).

\bibitem[{\citenamefont{Kuzyk}(2005)}]{kuzyk05.01}
\bibinfo{author}{\bibfnamefont{M.~G.} \bibnamefont{Kuzyk}},
  \bibinfo{journal}{Phys. Rev. Lett.} \textbf{\bibinfo{volume}{95}},
  \bibinfo{pages}{109402} (\bibinfo{year}{2005}).

\bibitem[{\citenamefont{P\'{e}rez~Moreno}(2004)}]{perez04.01}
\bibinfo{author}{\bibfnamefont{J.}~\bibnamefont{P\'{e}rez~Moreno}}, Ph.D.
  thesis, \bibinfo{school}{Washington State University} (\bibinfo{year}{2004}).

\bibitem[{\citenamefont{Champagne and Kirtman}(2005)}]{champ05.01}
\bibinfo{author}{\bibfnamefont{B.}~\bibnamefont{Champagne}} \bibnamefont{and}
  \bibinfo{author}{\bibfnamefont{B.}~\bibnamefont{Kirtman}},
  \bibinfo{journal}{Phys. Rev. Lett.} \textbf{\bibinfo{volume}{95}},
  \bibinfo{pages}{109401} (\bibinfo{year}{2005}).

\end{thebibliography}
\end{document}